\documentclass[10pt,reprint,prd,nofootinbib,superscriptaddress,twocolumn]{revtex4}

\usepackage{amsfonts}
\usepackage{amsmath}
\usepackage{amssymb}
\usepackage{mathrsfs}
\usepackage[english]{babel}
\usepackage{graphicx}
\usepackage{epsfig}
\usepackage{bm}
\usepackage{verbatim}
\usepackage[utf8]{inputenc}
\usepackage{booktabs}
\usepackage{multirow}
\usepackage{soul}
\usepackage{xcolor}
\usepackage[colorlinks=true,urlcolor=red,citecolor=red]{hyperref}
\usepackage[font=small]{caption}
\usepackage{float}
\usepackage{blindtext}
\usepackage{subfigure}
\usepackage{cancel}
\usepackage{csquotes}
\usepackage{verbatim}
\usepackage[titletoc,title]{appendix}
\usepackage{bbm}

\newcommand{\nn}{\nonumber}

\newcommand{\bc}{\begin{center}}
\newcommand{\ec}{\end{center}}

\newcommand{\mathsym}[1]{}

\definecolor{veronica}{rgb}{0.63, 0.36, 0.94}

\begin{document}

\title{A Novel Transfer Matrix Framework for Multiple Dirac Delta Potentials}
\author{Joaquín Figueroa}
\email{joaquin.figueroac@alumnos.uv.cl}
\affiliation{Instituto de {F}\'{i}sica y {A}stronom\'{i}a,~{U}niversidad de {V}alpara\'{i}so,~{A}venida~{G}ran~{B}reta\~{n}{a} $1111$,~Valpara\'{i}so,~{C}hile}
\author{Ivan Gonzalez}
\email{ivan.gonzalez@uv.cl}
\affiliation{Instituto de {F}\'{i}sica y {A}stronom\'{i}a,~{U}niversidad de {V}alpara\'{i}so,~{A}venida~{G}ran~{B}reta\~{n}{a} $1111$,~Valpara\'{i}so,~{C}hile}
\author{Daniel Salinas-Arizmendi}
\email{daniel.salinas@usm.cl}
\affiliation{Departamento de Física, Universidad T\'{e}cnica Federico Santa Mar\'{\i}a y Centro Científico-Tecnológico de Valparaíso, Casilla 110-V, Valpara\'{\i}so, Chile.}
\date{\today }

\begin{abstract}
We present an analytical framework for studying quantum tunneling through multiple Dirac delta potential barriers in one dimension. Using the transfer matrix method, we derive a closed-form expression for the total transfer matrix of a system composed of $N$ equally spaced delta barriers. 
In a systematic manner, a compact expression is obtained for the first element of the transfer matrix, based on triangular numbers.
This, in turn, allows us to compute the transmission coefficient exactly as a function of the number of barriers. 
The proposed method successfully reproduces well-known results for one and two barriers and efficiently captures complex interference effects for larger values of $N$, such as $N=4$.
\end{abstract}

\date{\today}
\maketitle

\section{Introduction}

The interaction of quantum particles with short-range potentials modeled by the Dirac delta distribution has become a widely studied topic due to its theoretical significance and versatility across various areas of physics. Quantum tunneling through multiple potential barriers is a fundamental problem in one-dimensional quantum mechanics, with applications ranging from semiconductor devices to quantum transport theory. Dirac delta functions are frequently employed to model ultra-thin barriers due to their mathematical simplicity and physical relevance. For instance, an infinite periodic array of delta potentials (the Kronig–Penney model) successfully reproduces the formation of electronic bands in crystals \cite{Erman:2018}, highlighting the utility of these potentials in the modeling of real materials. However, analyzing a finite number of delta barriers presents significant challenges due to quantum interference effects among multiple barriers. As the number of barriers increases, obtaining closed-form results becomes increasingly difficult. In particular, solving the boundary conditions for each barrier becomes intractable for large values of $N$ \cite{Erman:2018}. This difficulty underscores the need for an analytical expression for the total transfer matrix and the transmission probability in systems with multiple Dirac delta potential barriers. Such a solution would facilitate calculations and provide a deeper understanding of the transmission phenomenon in multi-barrier systems.

The problem of multiple barriers has been studied using various theoretical approaches. A common method is the transfer matrix approach \cite{Levi:2003},
in which $2\times 2$ matrices associated with each barrier and the intermediate regions are multiplied to obtain the total transfer matrix of the system. While this method is useful for a small number of barriers, it becomes highly complex for large values of $N$. To overcome these limitations, alternative approaches have been explored, such as the Lippmann–Schwinger equation \cite{Erman:2018}, which solves the scattering problem in an integral form. Specific cases with a few barriers have been analyzed in detail: for instance, the double delta barrier system ($N=2$) exhibits resonant transmission peaks \cite{CORDOURIER_MARURI_2011}. Moreover, configurations with three barriers have been studied to identify more complex quantum interference effects \cite{Cordourier:2014}. Additionally, finite periodic arrays of delta potentials (analogous to superlattice models in semiconductor materials) have been investigated, where phenomena such as perfect transmission resonances and threshold anomalies have been observed \cite{SAHU20094033,lapidus1982resonance,senn1988threshold}.

In this work, we develop an analytical formalism based on the transfer matrix method to systematically study the scattering and transmission of multiple Dirac delta barriers. We derive a closed-form expression for the total transfer matrix of a system of particles traversing $N$ equally spaced delta potentials, which allows us to construct an expression for the transmission probability, based on triangular numbers \cite{falcao2012note,caccao2023intrinsic}. 
By leveraging the properties of these triangular numbers, our method efficiently accounts for all interference contributions between the barriers and condenses the information into a compact expression for the first element of the transfer matrix. This allows us to obtain exact formulas for the transmission coefficient as a function of the number of barriers, providing a clearer understanding of how quantum transmission is affected by the system’s structure. Our approach significantly extends the analytical tools available for studying quantum tunneling in large $N$ systems.

\section{The Delta Dirac Potential}

\begin{figure}[]
\centering
\includegraphics[scale=0.4]{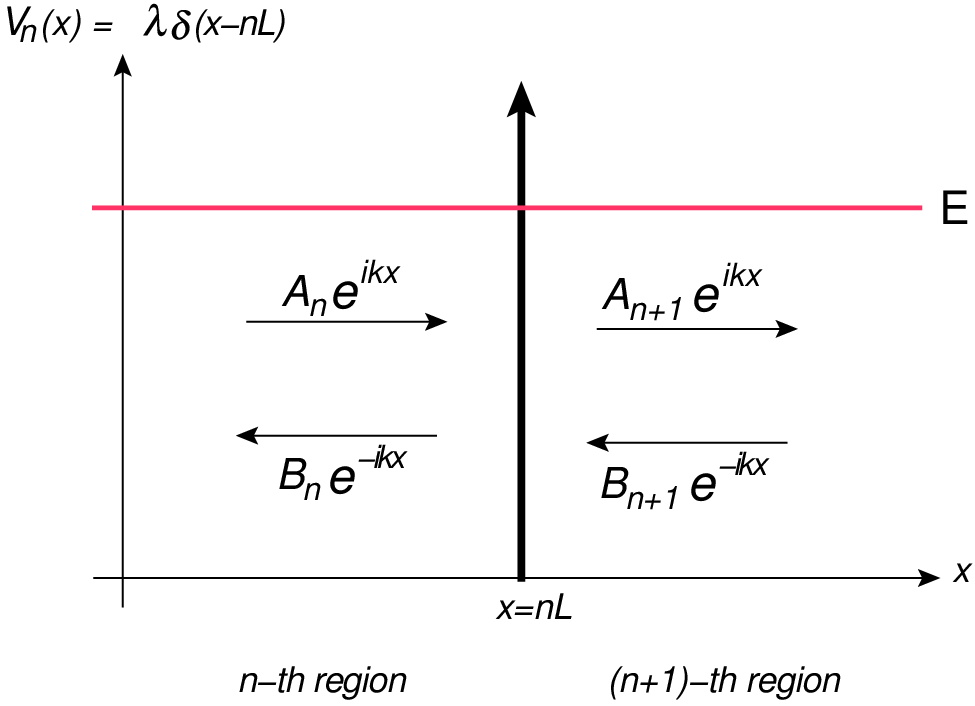}
\caption{Schematic representation of a one-dimensional delta potential 
located at $x = nL$. The diagram illustrates the incident particle with energy $E$ in the regions $n$ and $n+1$. The $n$-th potential has been drawn
with its corresponding incident ($A_n$), reflected ($B_n$) and transmitted ($A_{n+1}$) 
wave amplitudes, alongside with the reflected ($B_{n+1}$) wave amplitude of the $(n+1)$-th potential.}
\label{fig:delta1}
\end{figure}

When a particle interacts with a localized potential it is common to model potentials in terms of the Dirac delta function. 
For $N$-delta potentials, the potential can be written as:
\begin{equation} \label{sec2:eq:PotencialDeltas}
    V(x) = \lambda \sum_{n=1}^{N} \delta(x - nL),
\end{equation}
where $\lambda > 0$ denotes a potential barrier and $\lambda<0$ a potential well, and $nL$ $(n=1,...,N)$ denotes the distance of the $n$-th potential from the origin, as illustrated in Figure~\ref{fig:delta1}. In this section, we will focus on the scattering of a particle with energy $E$ and mass $m$ interacting with the $n$-th potential $V_n(x) =  \lambda \,\delta (x - nL) $. By solving the time-independent Schrödinger equation and imposing the appropriate boundary conditions, one can determine the corresponding transmission and reflection amplitudes of the wavefunction.

Let us begin with the time-independent Schr\"odinger equation for a particle interacting with the 
$n$-th
barrier ($\lambda > 0$), which is expressed as follows:
\begin{equation} \label{sec1:eq:Schrodinger}
-\frac{\hbar^2}{2m} \frac{d^2 \phi(x)}{dx^2} + V_n(x) \phi(x) = E \phi (x),
\end{equation}
where a particle incident from the left is partially reflected and partially transmitted by the potential barrier $V_n(x)$, for additional details, see Refs.~\cite{cohen1986quantum,messiah1999quantum}. The wavefunction in the $n$-th region is taken as
\begin{equation} \label{sec1:eq:FdeOnda}
    \phi_n (x) = A_{n} e^{ikx} + B_{n} e^{-ikx}
\end{equation}
This is valid for $(n-1)L\leq x \leq nL$, where $A_{n}$ and $B_{n}$ are the constant probability amplitudes and $k$ the wave number, so that $k^2=2mE/\hbar^2$. The wavefunction for the $(n+1)$-th region is taken as

\begin{equation} \label{sec1:eq:FdeOnda2}
    \phi_{n+1} (x) = A_{n+1} e^{ikx} + B_{n+1} e^{-ikx}
\end{equation}

which is valid for $nL \leq x \leq (n+1)L$, where $A_{n+1}$ and $B_{n+1}$ are the constant probability amplitudes for the $(n+1)$-th region.

This set of wavefunctions admits two boundary conditions. First, the set must be continuous 
at each point on the $x$-axis, which in our case occurs at $x=nL$ for regions 
$n$ and $n+1$. Hence,
\begin{equation} \label{sec1:eq:Continuidad}
    A_{n} e^{iknL} + B_{n} e^{-iknL} = A_{n+1} e^{iknL} + B_{n+1} e^{-iknL},
\end{equation}
while the second condition corresponds to the derivative’s discontinuity at 
$x=nL$, naturally induced by the delta potential. For two points infinitesimally 
close around $nL$, it can be written as
\begin{equation}\label{eq:discont1}
\begin{aligned}
      & -\frac{\hbar^2}{2m} \left [\left. \frac{d\phi}{dx} \right |_{x=nL+\epsilon} 
       - \left. \frac{d\phi}{dx} \right |_{x=nL-\epsilon} \right ] + \lambda 
       \phi(nL) \\
      &= E\int^{nL+\epsilon}_{nL-\epsilon} \phi(x) dx,
\end{aligned}
\end{equation}
and by taking the limit $\epsilon\to 0$, evaluated at the boundary, 
Eq.~\eqref{eq:discont1} becomes
\begin{equation}
    \frac{\hbar^2}{2m} \left [\left. \frac{d\phi_{n+1}}{dx} \right |_{x=nL} 
    - \left. \frac{d\phi_n}{dx}\right|_{x=nL} \right ] = \lambda \phi_n(nL),
\end{equation}
or equivalently,
\begin{equation} \label{sec1:eq:Discontinuidad}
\begin{aligned}
   (1+c) A_{n} e^{iknL} + (1-c) B_{n} e^{-iknL} = &  c A_{n+1} e^{iknL} \\
  & - c B_{n+1} e^{-iknL},   
\end{aligned}
\end{equation}
where $c$ is the parameter associated with the energy of the particles subject to scattering, defined by
\begin{equation}
 c=\frac{ik\hbar^2}{2m \lambda}   .
\end{equation}

The transfer matrix can be constructed from Eqs.~\eqref{sec1:eq:Continuidad} 
and \eqref{sec1:eq:Discontinuidad}. Making use of the simplifying definition 
\hbox{$K=e^{ikL}$}, we have
\begin{gather}
\begin{pmatrix}
K^n & K^{-n} \\
(1+c)K^n & (1-c)K^{-n} \\
\end{pmatrix}     
\begin{pmatrix}
A_n\\
B_n
\end{pmatrix} \\
\nn = 
\begin{pmatrix}
K^n & K^{-n} \\
cK^n & -cK^{-n} \\
\end{pmatrix}     
\begin{pmatrix}
A_{n+1}\\
B_{n+1}
\end{pmatrix}  ,
\end{gather}
or equivalently
\begin{equation}    
\begin{aligned}
 \begin{pmatrix}
A_n\\
B_n
\end{pmatrix} 
=  & 
\begin{pmatrix}
K^n & K^{-n} \\
(1+c)K^n & (1-c)K^{-n} \\
\end{pmatrix}^{-1}   \\
& \times \begin{pmatrix}
K^n & K^{-n} \\
cK^n & -cK^{-n} \\
\end{pmatrix}     
\begin{pmatrix}
A_{n+1}\\
B_{n+1}
\end{pmatrix} .
\end{aligned}
\end{equation}

Therefore, the relations for the amplitude coefficients in regions $n$ and $n+1$
are derived from a system in which the inverse transfer matrix is explicitly given by
\begin{equation}
\begin{pmatrix}
A_{n}\\
B_{n}
\end{pmatrix} 
=
\frac{1}{2c}
\begin{pmatrix}
2c-1 & -K^{-2n} \\
K^{2n} & 2c+1 \\
\end{pmatrix}     
\begin{pmatrix}
A_{n+1}\\
B_{n+1}
\end{pmatrix}.
\end{equation}
Note that this matrix implicitly depends on the position through the presence of the index $n$. We can eliminate the dependence on 
$n$ by applying the following normalization transformation:
\begin{equation} \label{eq:Coeficientes}
\begin{aligned}
\widetilde{A}_n= & \ A_n K^{n}, \\
\widetilde{B}_n= & \ B_n K^{-n}.
\end{aligned}
\end{equation}

Thus, the transfer matrix between two consecutive coefficients is given by:
\begin{equation}
\begin{pmatrix}
\widetilde{A}_{n}\\
\widetilde{B}_{n}
\end{pmatrix}
= \frac{\textbf{T}}{2c} \begin{pmatrix}
\widetilde{A}_{n+1}\\
\widetilde{B}_{n+1}
\end{pmatrix} ,
\end{equation}
where 
\begin{equation}
\textbf{T} = 
\begin{pmatrix}
(2c-1)K^{-1} & -K \\
K^{-1} & (2c+1)K \\
\end{pmatrix} 
\end{equation}

The above matrix is independent of $n$, \textit{i.e., it does not depend on the position of the potential barrier}. This property provides a significant advantage when generalizing an expression for $N$ potential barriers.

\section{$N$-Delta Dirac Potentials and the Principal Transfer Matrix}

\begin{figure}[]
\centering
\includegraphics[scale=0.5]{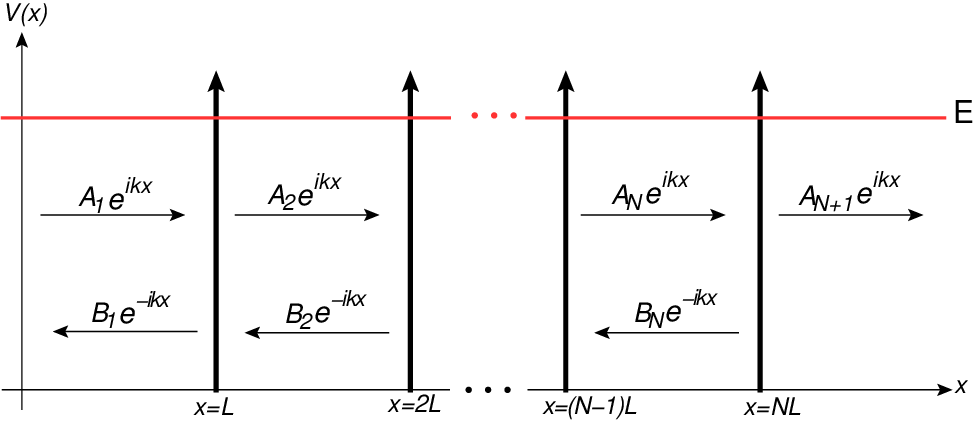}
\caption{Schematic representation of a one-dimensional arrangement with $N$ 
equally spaced delta potentials, located at $x = L, 2L, \ldots, NL$. 
The figure shows the incident wave amplitudes ($A_n e^{ikx}$) and reflected 
wave amplitudes ($B_n e^{-ikx}$) in each region, as well as the incident 
particle energy $E$.}
\label{fig:Ndelta}
\end{figure}

Now, we will consider the complete scattering process, taking into account the $N$ potential
barriers, Eq~\eqref{sec2:eq:PotencialDeltas}.
In Figure~\ref{fig:Ndelta}, we present the one-dimensional arrangement of an $N$-delta potential,
which consists of $N+1$ regions. Here, we consider a particle incident from the left, for which
the final reflection coefficient does not appear, \textit{i.e.}, $\widetilde{B}_{N+1} =0$,
given that no barrier is present in region $N+1$.

Our goal is to establish a relationship for the $2N+1$ amplitude coefficients.
For $N$ potential barriers, the continuity conditions and the derivative discontinuity from the preceding section remain unchanged.
Consequently, it is possible to build the corresponding relations through the transfer matrix approach.

To illustrate the procedure, let us first consider 
scenario of a $2$-delta potential (two delta barriers). Its transfer matrix for the first barrier is given by
\begin{equation}
\begin{pmatrix}
\widetilde{A}_{1}\\
\widetilde{B}_{1}
\end{pmatrix} 
= \frac{1}{2c}
\begin{pmatrix}
(2c-1)K^{-1} & -K \\
K^{-1} & (2c+1)K
\end{pmatrix}
\begin{pmatrix}
\widetilde{A}_{2}\\
\widetilde{B}_{2}
\end{pmatrix},
\end{equation}
and for the second barrier,
\begin{equation}
\begin{pmatrix}
\widetilde{A}_{2}\\
\widetilde{B}_{2}
\end{pmatrix} 
= \frac{1}{2c}
\begin{pmatrix}
(2c-1)K^{-1} & -K \\
K^{-1} & (2c+1)K
\end{pmatrix}
\begin{pmatrix}
\widetilde{A}_{3}\\
\widetilde{B}_{3}
\end{pmatrix}.
\end{equation}

By combining these two expressions, one obtains
\begin{equation}
\begin{pmatrix}
\widetilde{A}_{1}\\
\widetilde{B}_{1}
\end{pmatrix} 
= \frac{1}{4c^2}
\begin{pmatrix}
(2c-1)K^{-1} & -K \\
K^{-1} & (2c+1)K
\end{pmatrix}^2  
\begin{pmatrix}
\widetilde{A}_{3}\\
\widetilde{B}_{3}
\end{pmatrix},
\end{equation}
where $\widetilde{B}_{3}=0$.

By iterating this procedure step by step, one can derive the generalized relationship for an $N$-delta potential, which will play a key role in determining the transmission coefficients according to our novel approach. The final general expression takes the form:
\begin{equation} \label{sec2:eq:NDeltasMatrizFinal}
\begin{pmatrix}
\widetilde{A}_1\\
\widetilde{B}_1
\end{pmatrix} 
= \left(\frac{\textbf{T}}{2c} \right)^N
\begin{pmatrix}
\widetilde{A}_{N+1}\\
0
\end{pmatrix}.
\end{equation}

For convenience, we rewrite the \textbf{full inverse transfer matrix} in terms of a matrix $\mathbf{M}^{(N)}$, referred to as the 
\textbf{principal transfer matrix}, in the form
\begin{equation}
\mathbf{T}^{N} = \mathbf{M}^{(N)}.
\end{equation}

The transmission probability is defined as \hbox{$T_N = \left| A_{N+1} \right|^2 / \left| A_1 \right|^2$}, where the initial probability amplitude $A_{1}$ is written in terms of the transmitted amplitude at the region $N+1$ according to Eq.~\eqref{sec2:eq:NDeltasMatrizFinal}, taking into account Eq.~\eqref{eq:Coeficientes}, and the information relating the incident wave to the transmitted wave corresponds to the $(1,1)$ element of the matrix $\mathbf{M}^{(N)}$. The amplitude can be expressed as
\begin{equation}
\widetilde{A}_{1} = \frac{1}{\left( 2c\right) ^{N}} 
M
_{11}^{(N)}\,\widetilde{A}_{N+1}.
\end{equation}

Finally, the generalized expression for the transmission probability is:
\begin{equation}\label{eq:transmissionprobability}
T_N = \left(\frac{k \hbar^2}{\lambda m}\right)^{2N} \frac{1}{\left|M_{11}^{(N)}
\right|^2 }.
\end{equation}

We present a novel, compact result for the transmission probability,
which is valid in the regime of resonant transmission and perfect tunneling \cite{Cordourier:2014}.

\section{First Element of the Principal Transfer Matrix}
\label{sec:m11}

\begin{table}[]
    \centering
  \begin{tabular}{|l|l|}
\toprule[0.2mm]
\hline
\hline
$N$ & \multicolumn{1}{|c|}{$\textbf{M}_{11}
^{(N)}
(\alpha, \beta)$} \\
\hline $\mathbf{1}$ & $\alpha$ \\
$\mathbf{2}$ & $\alpha^2-1$ \\
$\mathbf{3}$ & $\alpha^3-(2 \alpha+\beta)$ \\
$\mathbf{4}$ & $\alpha^4-\left(3 \alpha^2+2 \alpha \beta+\beta^2\right)+1$ \\
$\mathbf{5}$ & $\alpha^5-\left(4 \alpha^3+3 \alpha^2 \beta+2 \alpha \beta^2+\beta^3\right)+(3 \alpha+2 \beta)$ \\
\hline
\hline \bottomrule[0.2mm]
\end{tabular}
    \caption{Some bivariate multinomials corresponding to the first element of the principal transfer matrix.
}
    \label{tab:m11}
\end{table}

\begin{table*}[htbp]
\renewcommand{\arraystretch}{1.8}
\centering
\begin{tabular}{|l|l|}
\toprule[0.2mm]
\hline
\hline $N$ & \multicolumn{1}{|c|}{$\textbf{M}_{11}
^{(N)}
(\alpha, \beta)$} \\ \hline
$\mathbf{1}$ & \multicolumn{1}{|l|}{$\alpha $} \\ 
$\mathbf{2}$ & \multicolumn{1}{|l|}{$\alpha ^{2}-$\fbox{$1$}} \\ 

$\mathbf{3}$ & \multicolumn{1}{|l|}{$\alpha ^{3}-$\fbox{$\left( 2\alpha +\beta\right) $}} \\ 
$\mathbf{4}$ & \multicolumn{1}{|l|}{$\alpha ^{4}-\fbox{$\left( 3\alpha^{2}+2\alpha \beta +\beta^{2}\right) $}+1$} \\
$\mathbf{5}$ & \multicolumn{1}{|l|}{$\alpha ^{5}-\fbox{$\left( 4\alpha^{3} +3\alpha^{2}\beta +2\alpha \beta^{2}+\beta^{3}\right) $}+\left( 3\alpha +2\beta \right) $} \\ 
$6$ & \multicolumn{1}{|l|}{$\alpha ^{6}-\fbox{$\left( 5\alpha^{4}+4\alpha^{3}\beta +3\alpha^{2}\mathbf{\beta 
}^{2}+2\alpha \beta^{3}+\beta^{4}\right) $}+\left(
6\alpha ^{2}+6\alpha \beta +3\beta ^{2}\right) -1$} \\ 
$\mathbf{7}$ & \multicolumn{1}{|l|}{$\alpha ^{7}-\fbox{$\left( 6\alpha^{5}+5\alpha^{4}\beta +4\alpha ^{3}\beta^{2}+3\alpha^{2}\beta^{3}+2\alpha \beta^{4}%
+\beta ^{5}\right) $}+\left( 10\alpha ^{3}+12\alpha ^{2}\beta
+9\alpha \beta ^{2}+4\beta ^{3}\right) -\left( 4\alpha +3\beta \right) $} \\ 
\hline 
\hline \bottomrule[0.2mm]
\end{tabular}
\caption{Bivariate multinomial structure of the first element of the principal transfer matrix, $M_{11}^{(N)}(\alpha, \beta)$. 
Each row corresponds to a different value of $N$, illustrating the recurrence relations in the coefficients. The second submultinomial coefficients are enclosed in boxes.}
\label{tab:m11_squard}
\end{table*}

In this section, we focus on the calculation of the first element $M_{11}^{(N)}$ of the principal transfer matrix $\textbf{M}^{(N)}$, which plays a crucial role in describing the scattering process in systems with multiple Dirac delta potentials. This matrix element can be expressed as a bivariate multinomial whose structure reveals recurrent patterns that can be described in terms of specific combinatorial sequences. In particular, it is shown that the coefficients of these multinomials are closely related to non-symmetric triangular numbers, which have been studied in the context of hypercomplex function theory and generalized Appell polynomials \cite{falcao2012note,caccao2023intrinsic}. These numbers arise from a one-parameter family of numerical triangles, whose properties include recurrence relations analogous to the Fibonacci sequence, thereby enabling a precise algebraic characterization of the terms in $M_{11}^{(N)}$.

Throughout this section, we will explicitly analyze the structure of $M_{11}^{(N)}$
for different values of $N$, identify the underlying regularities, and derive a general formula for this matrix element, providing a fundamental analytical tool for characterizing the transmission coefficients in quantum systems.

To begin, the principal transfer matrix can be explicitly expressed as:
\begin{equation}
    \mathbf{M}^{(N)} (\alpha ,\beta) 
    =
\left(\begin{array}{cc}
\alpha & -K \\
K^{-1} & \beta
\end{array}\right)^N,
\end{equation}
with 
\begin{equation}\label{eq:ab}
\begin{aligned}
\alpha = & \  (2c - 1) K^{-1},\\
\beta = & \ (2c + 1) K
\end{aligned}
\end{equation}
and  where the parameters satisfying $\alpha^* = -\beta$, $\alpha \beta = 4c^2 - 1$, $K^*=K^{-1}$ and $c^*=-c$.

Table \ref{tab:m11} shows the principal transfer matrix element, which can be expressed as a bivariate multinomial in the variables $\alpha$ and $\beta$ conveniently ordered, whose structure exhibits recurrent patterns that allow for generalization to any number $N$ of Dirac delta potentials.

The analysis of the first element of the principal transfer matrix, $M_{11}^{(N)}(\alpha, \beta)$, reveals recurrent structural patterns in the coefficients of its polynomial terms. For instance, for a given value of $N$, the first term of the multinomial takes the form $\alpha^N$. Specifically, the coefficients of the second submultinomial associated with $M_{11}^{(N)}(\alpha, \beta)$ for $N \geq 2$ are related to the sequence \textbf{OEIS-A004736} \cite{oeis}, which describes non-symmetric triangular numbers. These coefficients are explicitly shown in Table \ref{tab:m11_squard}, enclosed within boxes. This sequence, widely studied in the context of hypercomplex function theory and combinatorics, can be represented as $\{1\}, \{2,1\}, \{3,2,1\}, \{4,3,2,1\}, \{5,4,3,2,1\}, \{6,5,4,3,2,1\}$, where each row corresponds to a descending set starting from a positive integer $g \geq 0$, related to $N$ by $g = N - 2$. The second submultinomial exists starting from $N \ge 2$.

The structure of these coefficients can be described by the combinatorial generating formula
\begin{equation}
C_2(g, k) = \binom{1 + g - k}{1} \binom{k}{0}, \quad (0 \leq k \leq g),
\end{equation}
where $g$ is the degree of the submultinomial,  and $k$ identifies the $k$-th coefficient.
For instance, for $g = 3$ (or 
$N = 5$), the second submultinomial of $M_{11}^{(5)} (\alpha, \beta)$ is $4\alpha^3 + 3\alpha^2\beta + 2\alpha\beta^2 + \beta^3$, whose coefficients correspond to the already shown sequence $\{4, 3, 2, 1\}$, as highlighted in Table \ref{tab:m11_squard}. These patterns not only highlight the rich algebraic structure of the elements of $M_{11}^{(N)}(\alpha, \beta)$, but also provide an analytical tool for characterizing the transmission properties of quantum systems with multiple Dirac delta potentials.

Therefore, the second submultinomial can be expressed as
\begin{equation}
P_2^g(\alpha, \beta)=\sum_{k=0}^g C_2(g, k) \alpha^{g-k} \beta^k.
\end{equation}

\begin{table*}[htbp]
\renewcommand{\arraystretch}{1.8}
\centering
\begin{tabular}{|l|l|}
\toprule[0.2mm]
\hline
\hline
$N$ & \multicolumn{1}{|c|}{$\textbf{M}_{11}
^{(N)}
(\alpha, \beta)$} \\ \hline
$\mathbf{1}$ & \multicolumn{1}{|l|}{$\alpha $} \\ 
$\mathbf{2}$ & \multicolumn{1}{|l|}{$\alpha ^{2}-1$} \\ 
$\mathbf{3}$ & \multicolumn{1}{|l|}{$\alpha ^{3}-\left( 2\alpha +\beta
\right) $} \\ 
$\mathbf{4}$ & \multicolumn{1}{|l|}{$\alpha ^{4}-\left( 3\alpha ^{2}+2\alpha
\beta +\beta ^{2}\right) +$\fbox{$1$}} \\ 
$\mathbf{5}$ & \multicolumn{1}{|l|}{$\alpha ^{5}-\left( 4\alpha ^{3}+3\alpha
^{2}\beta +2\alpha \beta ^{2}+\beta ^{3}\right) +$\fbox{$\left(
3\alpha +2\beta \right) $}} \\ 
$\mathbf{6}$ & \multicolumn{1}{|l|}{$\alpha ^{6}-\left( 5\alpha ^{4}+4\alpha
^{3}\beta +3\alpha ^{2}\beta ^{2}+2\alpha \beta ^{3}+\beta ^{4}\right) +%
\fbox{$\left( 6\alpha^{2}+6\alpha \beta +3\beta %
^{2}\right) $}-1$} \\ 
$\mathbf{7}$ & \multicolumn{1}{|l|}{$\alpha ^{7}-\left( 6\alpha ^{5}+5\alpha
^{4}\beta +4\alpha ^{3}\beta ^{2}+3\alpha ^{2}\beta ^{3}+2\alpha \beta
^{4}+\beta ^{5}\right) +\fbox{$\left( 10\alpha ^{3}+12\alpha^{2}\beta +9\alpha \beta^{2}+4\beta %
^{3}\right) $}-\left( 4\alpha +3\beta \right) $} \\ 
\hline 
\hline \bottomrule[0.2mm]
\end{tabular}
\caption{
The first element of the principal transfer matrix, $M_{11}^{(N)}(\alpha, \beta)$,
where the third submultinomial coefficients are enclosed in boxes, emphasizing their structured pattern within the polynomial expansion.}
\label{tab:third_submultinomial}
\end{table*}

Similarly, for the third submultinomial, a comparable analysis can be performed for different values of $N$. The coefficients of the third submultinomial for $N \geq 4$ are related to the sequence \textbf{OEIS-A104633} \cite{oeis_a104633}, represented as $\{1\}$, $\{3, 2\}$, $\{6, 6, 3\}$, $\{10, 12, 9, 4\}$, $\cdots$, where each set corresponds to the coefficients of the third submultinomial for a given value of $N$. These coefficients are explicitly shown in Table \ref{tab:third_submultinomial}, enclosed within boxes.

This sequence describes a non-symmetric triangular pattern, which can be generated using the following combinatorial formula:
\begin{equation}
C_3(g, k) = \binom{2 + g - k}{2} \binom{1 + k}{1}, \quad (0 \leq k \leq g).
\end{equation}
with $g=N-4$.

For example, when $g = 3$ (equivalent to $N = 7$), the third submultinomial is $10\alpha^3 + 12\alpha^2\beta + 9\alpha\beta^2 + 4\beta^3$, whose coefficients correspond to $\{10, 12, 9, 4\}$.

Using the aforementioned relation, it is possible to derive the generating formula for the third submultinomial of degree $g$:
\begin{equation}
P_3^g(\alpha, \beta) = \sum_{k=0}^g C_3(g, k) \alpha^{g-k} \beta^k.
\end{equation}

Thus, the fourth submultinomial contains coefficients related to the sequence \textbf{OEIS-A103252} \cite{oeis_a103252} for $N \geq 6$. Generalizing, we find a structure that can be described as:
\begin{equation}
P_m^g(\alpha, \beta)=\sum_{k=0}^g C_m(g, k) \alpha^{g-k} \beta^k
\end{equation}
for the submultinomial of degree $g$ with $m > 0$, where the coefficient $C_m(g, k)$ is given by:
\begin{equation}
C_m(g, k)=\binom{m-1+g-k}{m-1}\binom{m-2+k}{m-2}.
\end{equation}

Finally, for the first element of the principal transfer matrix, we obtain the following analytical expression:
\begin{equation} \label{eq:solution1}
M_{11}^{(N)}(\alpha, \beta)=\sum_{n=0}^{\left\lfloor\frac{N}{2}\right\rfloor}(-1)^n P_{n+1}^{N-2 n}(\alpha, \beta),
\end{equation}
where $\left\lfloor\frac{N}{2}\right\rfloor$ corresponds to the integer part of $N/2$.

\section{Analizing some cases}

In this section, we apply the developed formalism to specific scattering scenarios involving multiple delta potentials. We begin with simpler configurations and gradually increase the complexity by introducing additional barriers. In particular, we show how the transfer matrix approach, combined with the general expression for the transmission probability, allows for a direct analysis of key parameters such as the barrier strength and the incident particle's energy. Moreover, results consistent with the wave-like interference and resonance effects that emerge as $N$ increases will be obtained. The straightforward application of these results is facilitated by the novel technique presented in this work.

\subsection{Case $N=1$}

As a first illustrative example, let us consider the simplest scenario of a single Dirac delta barrier located at $x=L$. In this case, the system comprises two regions (to the left and right of the barrier), and there is no further barrier in which the particle can be reflected after traversing the region $x > L$. 

Although, in a conventional approach, one would typically begin calculations by imposing the wavefunction continuity and the derivative discontinuity at $x=L$, our aim here is to show how the proposed formulation bypasses a step-by-step reintroduction of these procedures. Instead, we rely directly on the principal transfer matrix and its first element, leveraging the developments presented in the preceding sections.

Hence, the first element of the principal transfer matrix $\mathbf{M}^{(1)}\equiv \mathbf{T}$ is obtained directly from the technique described in the Sec.~\ref{sec:m11}. 
One finds that:
\begin{equation}
\label{eq:m11N1}
M_{11}^{(1)}(\alpha, \beta)=\alpha=(2 c-1) K^{-1}.
\end{equation}

By substituting the result of Eq.~\eqref{eq:m11N1} into the expression for the transmission probability (cf.\ Eq.~\eqref{eq:transmissionprobability}), the single delta barrier result is obtained 
\begin{equation}
T_1 =\frac{1}{1 + \left(\frac{m\lambda}{\hbar^2 k}\right)^2},
\end{equation}
this is in complete agreement with the standard result found by directly solving the boundary- and derivative-jump conditions at $x = L$ and confirms that the
rescaling in Eq.~\eqref{eq:Coeficientes} (which removes explicit dependence on $n$) does indeed recover the canonical single delta scattering expression.

\begin{figure}
\centering
\includegraphics[width=0.99\linewidth]{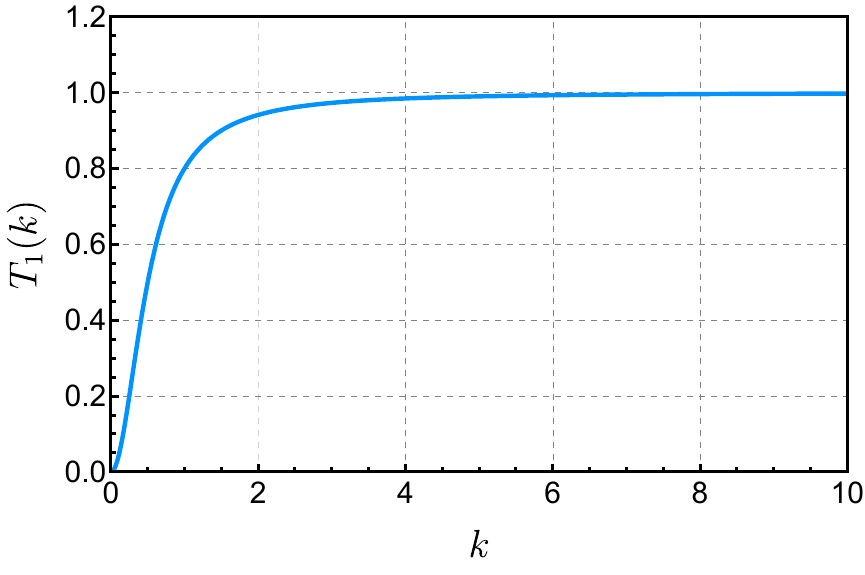}
\caption{Transmission probability $T_1$ for a single delta barrier ($N=1$) as a function of the wave number.}
\label{fig:t1}
\end{figure}

Numerically, the variation of the transmission probability $T_1$ (for $N=1$) as a function of the wave number $k$ is shown in Figure~\ref{fig:t1}, where we have set $\hbar^2 / (2m\,\lambda) = 1$.
One observes that for very small values of $k$, the particle has lower kinetic energy relative to the barrier height, resulting in a transmission $T_1$ 
significantly below unity. As $k$ increases, the delta potential becomes less relevant compared to the particle’s energy, and $T_1$ smoothly grows,  approaching $1$. This high-energy regime, where $k$ is large, corresponds to a situation in which the barrier is almost transparent and scattering effects are minimal.

\subsection{Cases $N=2$ and $N=4$}

Specifically for the case of $N \ge 2$ barriers, once the analytical expression for the first element of the transfer matrix has been obtained using Eq.~\eqref{eq:solution1}, it is convenient to present the result in the following form
\begin{equation}\label{eq:omega1}
M_{11}^{(N)}(\alpha,\beta) = \frac{\omega^{(N)}(c,K)}{K^{N}},
\end{equation}
where $\omega^{(N)}$ depends on the parameters $c$ and $K$, which themselves explicitly depend on the energy through the wave number parameter $k \sim \sqrt{E}$. Taking this dependence into consideration,the expression for the transmission probability becomes
\begin{equation}
T_N = \frac{4^N \left| c\right|^{2N}}{\left(\mathrm{Re}\ \omega^{(N)}(c,K)\right)^2+\left(\mathrm{Im}\ \omega^{(N)}(c,K)\right)^2}.
\end{equation}

This expression is useful for subsequently simplifying the cases presented in this section, as the new auxiliary variable in \eqref{eq:omega1} provides a simpler approach in the numerical study.

In the case of two delta barriers, the relevant information for determining the transmission corresponds to Eq.~\eqref{eq:solution1} for $N=2$, which is given by
\begin{equation}
M_{11}^{(2)}(\alpha,\beta) =\left( \alpha^2 - 1\right),
\end{equation}
and
\begin{equation}
\omega^{(2)}(c,K) = \left( 1- 2c\right)^2  -K^2.
\end{equation}

Subsequently, by combining the expression for $\omega^{(2)}(c,K)$ with the definition of $c$ and $K$, one arrives at the explicit form of the transmission coefficient,
written as
\begin{equation}
T_2 = \frac{ \left(\frac{\hbar^2 k}{m \lambda} \right)^4 
}{\left[1-\left(\frac{\hbar^2  k}{\lambda m}\right)^2-\cos (2 k L)\right]^2+\left[\frac{2 \hbar^2 k}{\lambda  m}+\sin (2 k L)\right]^2}.
\end{equation}

\begin{figure}
\centering
\includegraphics[width=0.99\linewidth]{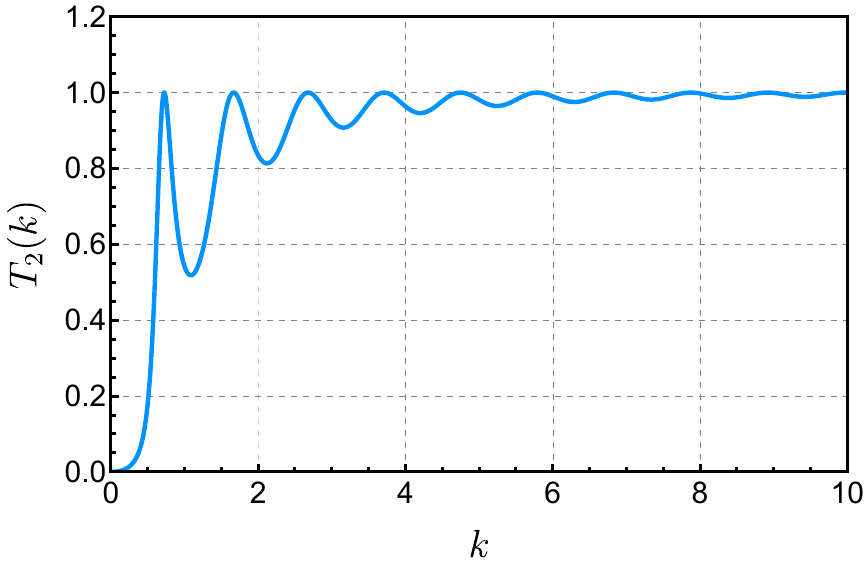}
\caption{Transmission probability $T_2$ for two delta barriers ($N=2$) as a function of $k$.}
\label{fig:t2}
\end{figure}

From the above expression for $T_2$, one can explicitly observe the dependence on both the particle's energy as well as on the separation $L$ and the intensity $\lambda$ of each delta barrier. In Figure~\ref{fig:t2} the dependence of the transmission probability $T_2$ on the wave number $k$ is shown, with $\hbar^2/(2m\lambda)=1$. In comparison with the single-barrier case, this behavior demonstrates how the reflected and transmitted waves at each barrier can overlap, producing interference patterns in the transmission probability. Moreover, the term $\cos(2 k L)$ highlights the importance of the phase acquired by the particle between the two barriers. The Eq.~\eqref{eq:solution1}, which gives $T_2$, may exhibit constructive interference arising from multiple scattering events, thereby contributing to the phenomenon of resonant tunneling; the conditions required for this study are presented in Ref.~\cite{CORDOURIER_MARURI_2011}.

Next, we consider the case of four delta barriers. This configuration leads to a considerably more complex interference pattern due to the increased number of scattering centers. Explicitly, the first element of the principal transfer matrix for $N=4$ can be expressed as
\begin{equation}
M_{11}^{(4)} \left(\alpha,\beta\right)= \alpha ^{4}-\left( 3\alpha^{2}+2\alpha \beta +\beta^{2}\right)+1, 
\end{equation}
and the where the auxiliary function $\omega^{(4)}$, is given by
\begin{equation}
\begin{aligned}
\omega^{(4)}(c,K) = & \  (1-2c)^4 -3(1-2c)^2K^2 + (3-8c^2)K^4 \\
& -(1+2c)^2 K^6   .
\end{aligned}
\end{equation}

\begin{figure}
\centering
\includegraphics[width=0.99\linewidth]{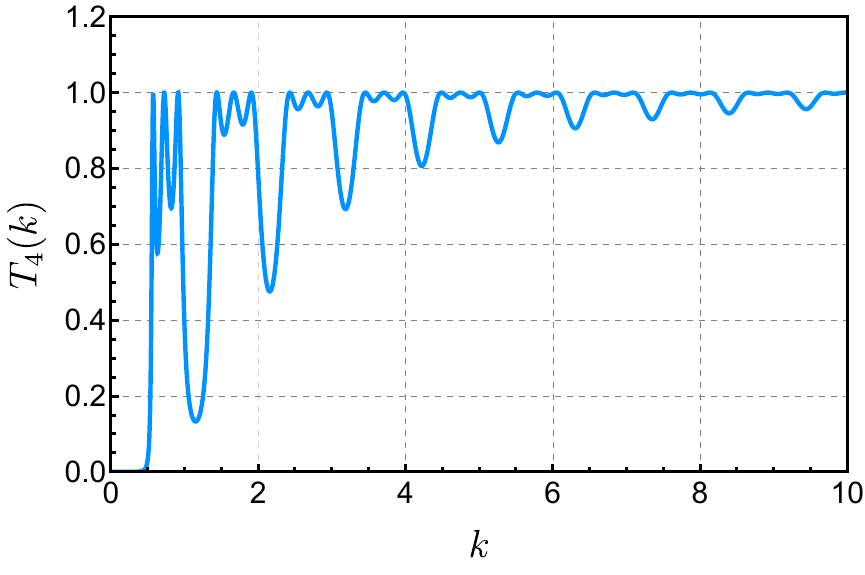}
\caption{Transmission probability $T_4$ for four delta barriers ($N=4$) as a function of $k$.}
\label{fig:t4}
\end{figure}

Thus, inserting the above expression for $\omega^{(4)}(c,K)$ into the general transmission probability formula, we obtain
\begin{equation}
T_4 = \frac{\left(\frac{k \hbar^2}{m \lambda} \right)^8}{\left(\mathrm{Re}\ \omega^{(4)}(c,K) \right)^2+\left(\mathrm{Im}\ \omega^{(4)}(c,K) \right)^2},
\end{equation}
where the real and imaginary part of $\omega^{(4)}(c,K)$ is
\begin{widetext}
\begin{equation}
\begin{aligned}
\mathrm{Re}\ \omega^{(4)}(k)= &\ 1-6\left(\frac{\hbar^2 k}{m \lambda}\right)^2+\left(\frac{\hbar^2 k}{m \lambda}\right)^4 + \left(3+2\left(\frac{\hbar^2 k}{m \lambda}\right)^2\right)\cos(4kL)+2\frac{\hbar^2 k}{m \lambda}\sin(6kL)\\
&-3\left[\left(1-\left(\frac{\hbar^2 k}{m \lambda}\right)^2\right)\cos(2kL)+2\frac{\hbar^2 k}{m \lambda}\sin(2kL)\right]-\left(1-\left(\frac{\hbar^2 k}{m \lambda}\right)^2\right)\cos(6kL)
.
\end{aligned}
\end{equation}

\begin{equation}
\begin{aligned}
\mathrm{Im}\ \omega^{(4)}(k)=&-4\frac{\hbar^2 k}{m \lambda}\left(1-\left(\frac{\hbar^2 k}{m \lambda}\right)^2\right) + \left(3+2\left(\frac{\hbar^2 k}{m \lambda}\right)^2\right)\sin(4kL)-3\left(1-\left(\frac{\hbar^2 k}{m \lambda}\right)^2\right)\sin(2kL)\\
&
+6\frac{\hbar^2 k}{m \lambda}\cos(2kL)-\left(1-\left(\frac{\hbar^2 k}{m \lambda}\right)^2\right)\sin(6kL)
-2\frac{\hbar^2 k}{m \lambda}\cos(6kL).
\end{aligned}
\end{equation}

\end{widetext}
In Figure~\ref{fig:t4}, the transmission probability $T_4$ for four delta barriers is shown as a function of the wave number, exhibiting a highly complex oscillatory behavior characterized by near-unit peaks that reveal perfect tunneling resonances. The presence of harmonic terms indicates that constructive and destructive interference among the delta barriers is extremely sensitive to the incident particle’s energy, the potential intensities, and their spacing. This resonant pattern, interpreted as the formation of quasi-bound states within the system, is consistent with previous studies on delta-potential arrays, where recurrence in transmission is linked to phase accumulation and the effective reconfiguration of the potential as dictated by the geometry of the arrangement \cite{Cordourier:2014,Ciccarello_2006}.

\subsection{Cases of large $N$ values}

For large values of $N$, the analytical determination of the principal element of the transfer matrix $M_{11}^{(N)}(\alpha, \beta)$ becomes increasingly intricate due to the combinatorial nature of its coefficients. Nevertheless, the systematic approach previously introduced in this manuscript remains effective and practical even in the regime of large $N$. Specifically, we begin by applying the general analytical formula provided by Eq.~\eqref{eq:solution1} to obtain $M_{11}^{(N)}(\alpha, \beta)$. Subsequently, we simplify the resulting expression by utilizing the auxiliary variable $\omega^{(N)}(c,K)$ defined in Eq.~\eqref{eq:omega1}. This procedure significantly reduces the algebraic complexity and facilitates numerical computations for the transmission probability $T_N$.

As $N$ increases, the transmission probability exhibits increasingly dense resonance structures due to the effects of multiple scattering and interference within the system. Such resonances correspond to quasi-bound states that form between consecutive barriers, and their spacing decreases as the number of barriers increases, eventually forming a quasi-continuous spectrum of resonant states. This phenomenon underscores the importance of accurately capturing interference effects when analyzing quantum transport properties in structures with multiple barriers. Therefore, having a method available to derive analytical expressions for large $N$ is relevant to various branches of physics that converge on this topic.

In Table~\ref{tab:third_submultinomial}, we present the results for $M_{11}^{(N)}(\alpha, \beta)$ for $N<8$ obtained via our method, thereby demonstrating its practicality and validating its applicability to configurations with an even greater number of barriers. The intention is that these results will, in the future, provide analytical expressions to study tunneling phenomena, asymptotic behavior, and other processes related to the scattering of particles by Dirac delta potential barriers.

\section{Conclusion}

In this work, we have developed an analytical approach to the problem of quantum tunneling through multiple Dirac delta barriers in one dimension. Using the transfer matrix method, we derived a closed-form expression for the total transfer matrix of a system composed of $N$ equally spaced delta potentials. Our approach, based on triangular numbers, allowed us to efficiently account for all interference contributions between the barriers, obtaining a compact expression for the first element of the transfer matrix. This, in turn, enabled us to derive an exact formula for the transmission coefficient as a function of the number of barriers.

The results obtained confirm the validity and efficiency of our method, reproducing well-known cases such as the scenarios of one and two delta barriers. Moreover, for large values of $N$, such as $N=4$, our analytical framework successfully captures complex interference effects. Therefore, the analytical
tools presented in this manuscript provide a versatile and systematic framework suitable for studying and designing quantum devices based on multiple Dirac delta potentials in the limit of large $N$. Future studies could further investigate asymptotic behaviors and scaling properties in this regime, thereby offering additional insights into quantum tunneling phenomena in complex potential landscapes.

\section*{Acknowledgments}
This research has received funding from Fondecyt (Chile), Grant 
No~1210131, ANID PIA/APOYO AFB230003.  I.G.
would like to thank CEFITEV-UV for partial support.

\bibliographystyle{utphys}
\bibliography{ref}
\end{document}